\documentclass[aps,twocolumn,showpacs,nofootinbib]{revtex4}
\usepackage[english]{babel}
\usepackage{amsmath,amssymb,graphicx,hyperref,natbib}

\catcode`\<=\active \def<{
\fontencoding{T1}\selectfont\symbol{60}\fontencoding{\encodingdefault}}
\newcommand{\cdummy}{\cdot}
\newcommand{\nobracket}{}
\newcommand{\tmem}[1]{{\em #1\/}}
\newcommand{\tmop}[1]{\ensuremath{\operatorname{#1}}}
\newcommand{\tmtextit}[1]{{\itshape{#1}}}

\begin{document}

\title{Topography-induced persistence of atmospheric patterns}

\author{D. Ciro, B. Raphaldini, C. Raupp}
\affiliation{Institute for Astronomy, Geophysics and Atmospheric Sciences\\
University of Sao Paulo}

\begin{abstract}
  Atmospheric blockings are persistent large-scale climate patterns with
  duration between days and weeks. In principle, blockings might involve a
  large number of modes interacting non-linearly, and a conclusive description
  for their onset and duration is still elusive. In this paper we introduce a
  simplified account for this phenomena by means of a single-triad of
  Rossby-Hawritz waves perturbed by one topography mode. It is shown that the
  dynamical features of persistent atmospheric patterns have {\tmem{zero}}
  measure in the phase space of an unperturbed triad, but such measure becomes
  finite for the perturbed dynamics. By this account we suggest that static
  inhomogeneities in the two-dimensional atmospheric layer are required for
  locking flow patterns in real space.
\end{abstract}

{\maketitle}

\section{Introduction}\label{sec.introduction}

Atmospheric blocking is a phenomena whereby the basic predominantly zonal flow
of the atmosphere breaks, acquiring a significant meridional component,
possibly correlated with inhomogeneities such as topography and thermal
forcing, e.g. by ocean/continent contrasts. Although time fraction of blocked
states of the atmosphere is relatively small, estimated between 2\% and 22\%
in the Northern Hemisphere {\cite{HamillKiladis2014}}, it is enough to
influence the climatological mean and result in weather conditions that depart
from the mean and are often extreme, causing important socio-economic impacts.
Blocking phenomena effects are most strongly felt in the Northern Hemisphere
in mid-latitudes where blocked states result in advection of cold winds coming
from the Arctic Pole, causing episodes of extreme cold during the winter
{\cite{SillmannEtAl2011}}, {\cite{PfahlWernli2012}}, although less
significant, blocking events occur in the Southern hemisphere
{\cite{DeLimaAbrizzi2002}}, {\cite{Renwick1998}}.

Proper representation of the onset and duration of blocking events still pose
difficulties in climate modeling and weather forecasts, often limiting the
predictability {\cite{HamillKiladis2014}}, and being underestimated in
frequency of occurrence {\cite{ScaifeEtAl2011}}. Further research on the basic
mechanisms behind blocking transitions, may lead to a better understanding and
improvements in its predictability in the future.

Theoretical models of these transitions usually rely on barotropic
quasi-geostrophic equation with inclusion of topography and/of inhomogeneus
thermal forcing, models can be either forced-dissipative or Hamiltonian. In the first scenario zonal and different blocked states constitute
equilibria of the underlying dynamical system, these equilibria usually
can be characterized as selective decay states of the equations which are
zonal flows and flows correlated with the topography
{\citealp{BrethertonHaidvogel1976}}, therefore transitions between zonal and
blocked states can be understood as trajectories of the system connecting the
vicinities of different equilibria in phase space.

In this paper we are interested in the conservative dynamics of the system, so
that viscosity will be neglected and the system becomes Hamiltonian. From the
point of view of wave turbulence, transitions between flow patterns are the
result of the nonlinear interaction between Rossby waves, where
there is a zonal flow is represented by a Rossby mode with {\tmem{zero}} zonal
wave number and therefore {\tmem{zero}} frequency, and other modes with
meridional flows and nonzero frequency, that absorb its energy.

Modal decomposition of the barotropic quasi-geostrophic equation in Rossby
waves lead to the existence of wave triads, where three modes exchange energy
non-linearly, and each mode can be member of different triads simultaneously.
This picture, allow us to interpret the complex behavior of the full system in
terms the energy flow between triads which are the non-linear units of the
full dynamical system.

In this paper, we investigate in detail the dynamics of a single triad of the
barotropic quasi-geostrophic equation with and without topography. In
particular, we are interested in the existence of triads developing
high-amplitude, non-drifting Rossby waves, in correspondence to persistent
blocking patterns in the atmosphere. It is shown that in the simple case
without topography such states exist amid periodic solutions of the triad
mismatch phase, but the solution has {\tmem{zero}} measure in the solution
space, i.e. it is statistically impossible. However, when introducing one
congruent topography mode as a small perturbation to the system, the measure
of such solutions becomes finite, and high-amplitude non-drifting states
become available for a range of initial conditions in phase space. This
important feature emerges from important topological changes in the invariants
of the perturbed problem, illustrating the critical role of topography in this
simplified account of the phenomena.

The paper is organized as follows. In Sec. \ref{sec.model} we introduce the
Rossby waves decomposition of barotropic quasi-geostrophic model with
topography, then, in Sec. \ref{sec.single_triad} the dynamics of a single
triad is studied in detail and the {\tmem{zero}}-drift solutions are
characterized. In Sec. \ref{sec.triad_topography} the topography is introduced
and its effects on the {\tmem{zero}}-drift solutions is numerically
investigated. In Sec. \ref{sec.conclusions} we conclude the main content with
our conclusions and perspectives. In the Appendix section \ref{sec.appendix}
relevant calculations and approximations are presented.

\section{The model}\label{sec.model}

In this work we consider a simplified barotropic quasi-geostrophic model of
the atmosphere with variable depth, between the topography and the free upper
surface. By decomposing the depth in a constant and a fluctuating part $H =
\overline{H} + \tilde{H}$, and by means of a perturbative expansion described
in the Appendix \ref{sec.ap_model_derivation}, it can be shown that the
relative vorticity satisfies
\begin{equation}
  \frac{\partial \zeta}{\partial t} + J_a (\psi, \zeta + f - h),
  \label{eq.barotropic_vorticity_topography}
\end{equation}
where $\psi$ is the velocity stream, $\zeta = \nabla^2 \psi$ is the relative
vorticity, $f = 2 \Omega \cos \theta$ is the planetary vorticity or the
Coriolis parameter and $h = f \tilde{H} / \bar{H}$ is the depth fluctuation
modulated by the Coriolis parameter in units of its mean value in the
atmospheric depth. This fluctuation is due to both the topography and the free
surface, but here we will refer to it as the topography. In the following, the
Jacobian function $J$ is defined in terms of regular spherical coordinates
with $\theta = 0$ in the north pole
\[ J_a (g_1, g_2) = \frac{1}{a^2 \sin \theta} \left( \frac{\partial
   g_1}{\partial \theta} \frac{\partial g_2}{d \phi} - \frac{\partial
   g_1}{\partial \phi} \frac{\partial g_2}{d \theta} \right), \]
where $a$ is the mean planetary radius. It is worth noting that during the
derivation of this model we did not employed the beta-plane approximation, so
that $\psi$ is the global velocity stream, and the velocity can be obtained
anywhere via
\[ \vec{u} = (1 - H / \overline{H})  \hat{n} \times \nabla \psi \]
where $\hat{n}$ is the unit vector normal to the planetary surface and
$\vec{u}$ is a non-solenoidal field in contrast to the usual barotropic
vorticity models.

To perform a more universal analysis we measure the velocities in units of the
characteristic flow velocity $u_0$, the distances in units the mean planetary
radius $a$, and the time in units of $(2 \Omega)^{- 1}$, leading to a
dimensionless version of (\ref{eq.barotropic_vorticity_topography})
\begin{equation}
  \frac{\partial \zeta}{\partial t} + \frac{\partial \psi}{\partial \phi} +
  J_1 (\psi, \lambda \zeta - \varepsilon h) = 0,
  \label{eq.dimensionless_model}
\end{equation}
where $\lambda = u_0 / 2 \Omega a$ is the ratio between the characteristic
flow velocity and the equatorial planetary velocity, the stream $\psi$ is
measured in units of $u_0 a$, $\zeta$ in units of $u_0 / a$, and the Jacobian
operator is taken in a unit sphere. The parameter $\varepsilon = | \tilde{H}
|_{\max} / \overline{H}$, is the maximum topography value, so that the
relative topography $h (\theta, \phi)$ is of order {\tmem{one}}, as well as
$\psi (\theta, \phi)$ and $\zeta (\theta, \phi)$.

Since $\lambda$ and $\varepsilon$ are small parameters and all the fields are
of the same order the dominant role of the Rossby-Hawritz waves becomes
explicit and the nonlinear terms in the Jacobian provide small corrections to
the evolution equation that enable energy transfers between modes. Due to the
completeness of the spherical harmonics we can expand the stream function
$\psi$ and the depth $h$ as Laplace's series,
\begin{eqnarray}
  \psi (\theta, \phi, t) & = & \sum_{m, n} a_{m, n} (t) Y_n^m (\theta, \phi), 
  \label{eq.stream_expansion}\\
  h (\theta, \phi) & = & \sum_{m, n} h_{m, n} Y_n^m (\theta, \phi), 
  \label{eq.topography_expansion}
\end{eqnarray}
where, in the linear regime the stream amplitudes oscillate with the
Hawritz-Rossby frequencies (here re-scaled by $2 \Omega$).
\begin{equation}
  \omega_{m, n} = \frac{m}{n (n + 1)} = \frac{m}{k_n} .
  \label{eq.hawritz-rosby_freq}
\end{equation}
Plugging (\ref{eq.stream_expansion}) and (\ref{eq.topography_expansion}) in
(\ref{eq.dimensionless_model}) and using the orthogonality of the spherical
harmonics together with well known selection rules we can derive a reduced
dynamical system involving various amplitudes $a_i = a_{m_i, n_i}$, which can
be organized in {\tmem{triads}} $\{ \psi_{\alpha}, \psi_{\beta}, \psi_{\gamma}
\}$, satisfying $m_{\alpha} + m_{\beta} = m_{\gamma}$. In general, triads are
coupled by shared modes that transfer energy between them, and the energy of
the system can flow in complicated patterns in the modal space.

\subsection{Equations for a single triad with
topography}\label{sec.single_triad}

Instead of considering complex modal structures, we focus our attention in the
dynamics of a single triad with topography, which sets the basic development
scenarios that get modified by the interaction with other modes. The
construction of the dynamical system and the equations for the general problem
with unspecified modal structure can be found in Appendix
\ref{sec.ap_reduced_system}.

For a single triad $\{ \psi_1, \psi_2, \psi_3 \}$, and its conjugated modes,
with one single topography mode compatible with $\psi_3$ the amplitudes evolve
according to
\begin{eqnarray*}
  i n_1 (n_1 + 1) \dot{a}_1 & = & - m_1 a_1 + \lambda K \mu_{32} a_2^{\ast}
  a_3 + \varepsilon K h a_2^{\ast},\\
  i n_2 (n_2 + 1) \dot{a}_2 & = & - m_2 a_2 + \lambda K \mu_{13} a_1^{\ast}
  a_3 - \varepsilon K h a_1^{\ast},\\
  i n_3 (n_3 + 1) \dot{a}_3 & = & - m_3 a_3 + \lambda K \mu_{12} a_1 a_2,
\end{eqnarray*}
where $\mu_{\beta \alpha} = k_{\beta} - k_{\alpha}$, and $K$ is the
interaction coefficients between the modes as described in Appendix
\ref{sec.ap_reduced_system}. By requiring the stream function to be
real-valued we obtain the form
\begin{equation}
  \psi (\theta, \phi, t) = \sum_{i = 1}^3 | a_i | P_{n_i}^{m_i} (\cos \theta)
  \cos [m_i \phi + \beta_i (t)],
\end{equation}
where $\beta_i (t)$ is the phase of the complex variable $a_i$. In its present
form the system can be easily fed with physical parameters, but more useful
predictions can be made if we re-scale the variables to identify a more
fundamental combination of parameters that allows to compare all triads in the
same footing.

To to this we define the constants
\[ \alpha_1 = k_2^{- 1} - k_3^{- 1}, \quad \alpha_2 = k_3^{- 1} - k_1^{- 1},
   \quad \alpha_3 = k_2^{- 1} - k_1^{- 1}, \]
and a new set of dynamical variables
\[ z_i = \lambda K (\alpha_j \alpha_k)^{1 / 2} k_i a_i e^{i \varphi_{0, i}},
   \quad i \neq j \neq k. \]
where the phases $\varphi_{0, i}$ are yet to be determined, and satisfy
$\varphi_{0, 3} = 2 \varphi_{0, 1}$, $\varphi_{0, 1} = \varphi_{0, 2}$. Then,
we put the topography amplitude in its polar form $h = | h | e^{i \varphi_h}$,
and introduce two dependent perturbation amplitudes
\begin{eqnarray}
  \varepsilon_1 & = & \frac{k_3}{k_3 - k_1} K \sqrt{| \alpha_1 \alpha_2 | }  |
  h | \varepsilon, \\
  \varepsilon_2 & = & \frac{k_3}{k_3 - k_2} K \sqrt{| \alpha_1 \alpha_2 | }  |
  h | \varepsilon, 
\end{eqnarray}
then, the phase of $\varepsilon h$ becomes $\varphi_h + \beta_{12}$, where
$\beta_{12} = \pi / 2$ if $\tmop{sgn} (\alpha_1 \alpha_2) = - 1$, and
$\beta_{12} = 0$ otherwise. To express the system in terms of universal
parameters alone we can set $\varphi_{0, 1} = - \varphi_h - \beta_{12}$, which
fixes all phases and leads to the most compact form for this dynamical system
\begin{eqnarray}
  i \dot{z}_1 & = & - \omega_1 z_1 + z_2^{\ast} z_3 + \varepsilon_1
  z_2^{\ast}, \nonumber\\
  i \dot{z}_2 & = & - \omega_2 z_2 + z_1^{\ast} z_3 + \varepsilon_2
  z_1^{\ast},  \label{eq.universal_triad}\\
  i \dot{z}_3 & = & - \omega_3 z_3 + z_1 z_2 . \nonumber
\end{eqnarray}
where $\varepsilon_{1, 2}$ are real and can be positive or negative and
satisfy $\varepsilon_1 / \varepsilon_2 = (k_3 - k_2) / (k_3 - k_1)$. In
general, the amplitudes $z_i (t)$ are complex variables and the previous
dynamical system is {\tmem{six}}-dimensional and non-linear, but it can be
tackled analytically in a number of useful particular situations. Instead of
studying numerically this complex system we proceed analytically from the less
general single triad without topography to the more complete situation of two
triads with zonal flow and topography. This will allow us to identify the most
relevant features of the system and the variables that contain most relevant
information of the system. The numerical analysis will only be employed as an
illustrative tool after we identify the relevant features of the system.

\section{single triad dynamics}

The dynamics of a single triad without topography is completely integrable,
i.e. there are sufficient constants of motion to reduce the dynamics to two
dimensions, where systems are in general integrable. Taking $\varepsilon_1 =
\varepsilon_2 = 0$ in (\ref{eq.universal_triad}) the dynamical system can be
derived from the Hamilton equations $\dot{z_i} = \partial \mathcal{H}/
\partial z_i^{\ast}$, and $\dot{z}_i^{\ast} = - \partial \mathcal{H}/ \partial
z_i$, with Hamiltonian function
\begin{eqnarray}
  \mathcal{H} (\{ z_i \}, \{ z_i^{\ast} \}) & = & i (\omega_1 | z_1 |^2 +
  \omega_2 | z_2 |^2 + \omega_3 | z_3 |^2 \nobracket \nonumber\\
  &  & - \nobracket z_1 z_2 z_3^{\ast} - z_1^{\ast} z_2^{\ast} z_3), 
  \label{eq.hamiltonian}
\end{eqnarray}
this will be relevant below for the construction of a complete set of
constants of motion.

Writing the complex amplitudes in polar form $z_i (t) = r_i (t) e^{i \varphi_i
(t)}$ we obtain a real variables dynamical system
\begin{eqnarray}
  \dot{r_1} & = & r_2 r_3 \sin \varphi,  \label{eq.dotr1}\\
  \dot{r_2} & = & r_1 r_3 \sin \varphi,  \label{eq.dotr2}\\
  \dot{r_3} & = & - r_1 r_2 \sin \varphi,  \label{eq.dotr3}\\
  \dot{\varphi} & = & \omega_{} + \left( \frac{r_2 r_3}{r_1} + \frac{r_1
  r_3}{r_2} - \frac{r_1 r_2}{r_3} \right) \cos \varphi,  \label{eq.dotphi}
\end{eqnarray}
where $\omega = \omega_3 - \omega_1 - \omega_2$, and $\varphi = \varphi_3 -
\varphi_1 - \varphi_2$, are the frequency and phase mismatch respectively. In
the unperturbed situation, the polar representation already reduces the
dimension of the problem to {\tmem{four}}.

By combining (\ref{eq.dotr1}, \ref{eq.dotr2}) and (\ref{eq.dotphi}) we can
find two constants of motion
\begin{eqnarray}
  I^2_1 & = & r_1^2 + r_3^2,  \label{eq.I1}\\
  I^2_2 & = & r_2^2 + r_3^2,  \label{eq.I2}
\end{eqnarray}
which represent circles of radius $I_1$ and $I_2$ in the spaces $r_1 - r_3$
and $r_2 - r_3$ respectively, allowing us to put $r_1$ and $r_2$ in terms of
$I_1, I_2$ and $r_3$ alone, reducing the dimension of the system to
{\tmem{two.}} Provided that $r_3 < I_{1, 2}$ at all times, we can make more
universal observations by measuring $r_3$ in units of $I_{\min} = \min (I_1,
I_2)$, so that the resulting {\tmem{two}}-dimensional dynamical system takes
the form
\begin{eqnarray}
  \frac{d x}{d \tau} & = & - (1 - x^2)^{1 / 2}  (\kappa^2 - x^2)^{1 / 2} \sin
  \varphi,  \label{eq.2d_system_x}\\
  \frac{d \varphi}{d \tau} & = & \overline{\omega} + \left( \frac{x}{\kappa^2
  - x^2} + \frac{x}{1 - x^2} - \frac{1}{x} \right) \times \nonumber\\
  &  & (1 - x^2)^{1 / 2}  (\kappa^2 - x^2)^{1 / 2} \cos \varphi 
  \label{eq.2d_system_phi}
\end{eqnarray}
where the amplitude and time were re-scaled by $x = r_3 / I_{\min}$, $\tau =
I_{\min} t$, and there are only two parameters to control the dynamics $\kappa
= I_{\max} / I_{\min}$, $\overline{\omega} = \omega / I_{\min}$.

This system encodes all the dynamics of a single triad for given values of the
mismatch $\omega$, and constants $I_1, I_2$, determined by the initial
amplitudes of the modes. In principle, these equations can be integrated to
determine $r_3$ as a function of time, but the geometry of phase space $x -
\varphi$, contains a great deal of information without requiring to integrate
the dynamical system. To exploit this fact we just need to determine the fixed
points of (\ref{eq.2d_system_x}-\ref{eq.2d_system_phi}) and their type of
stability. In Appendix \ref{sec.ap_fixed_points} this is done with some
detail, resulting in six different equilibria, two pairs of hyperbolic points
or saddles, one pair $S_0^{\pm}$ at $(\varphi, x) = (\pm \pi / 2, 0)$ and the
other \ $S_0^{\pm}$, at $(\varphi, x) = (\pm \pi / 2, 1)$, and two elliptic
points or centers $C_0, C_{\pi}$ at $\varphi = 0$ and $\pi$, for which the
corresponding values of $x$ can be determined numerically from high order
polynomial equation (see Appendix \ref{sec.ap_fixed_points}).

\begin{figure}[h]
  \resizebox{210pt}{170pt}{\includegraphics{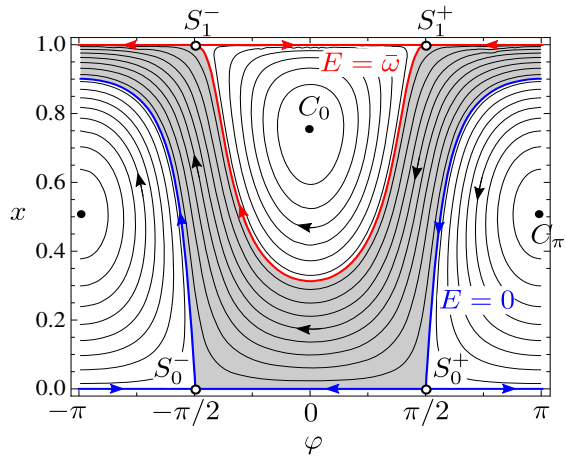}}
  \caption{\label{fig.contours_triad}Contours of (\ref{eq.const_E}) and fixed
  points of (\ref{eq.2d_system_x}-\ref{eq.2d_system_phi}) for a triad with
  $\overline{\omega} = - 0.76$, and $\kappa = 1.2$. The phase space is
  separated in regions with bounded (white) and unbounded motions (gray) in
  $\varphi$. The separatrices in blue and red contain solutions with infinite
  period and delimit the regions with finite period solutions.}
\end{figure}

Let us choose for illustrative purposes a triad with rescaled mismatch
$\overline{\omega} = - 0.76$, and $\kappa = 1.2$. in Fig.
\ref{fig.contours_triad}, we show the fixed points, and how they organize the
phase-space flow, mainly, the saddle points, which are connected to the
separatrices that delimit the regions with oscillations and rotations in
$\varphi$. To put $x$ in terms of $\varphi$ alone we can use the fact that the
Hamiltonian function (\ref{eq.hamiltonian}) is explicitly independent of time,
and consequently a constant of motion. Putting (\ref{eq.hamiltonian}) in polar
form we can derive a real constant of motion that couples $r_3$ and $\varphi$
in terms of the other system parameters
\begin{eqnarray*}
  \tmop{Im} \mathcal{H} (r_3, \varphi) & = & \omega_1 r_1^2 (r_3) + \omega_2
  r_2^2 (r_3) + \omega_3 r_3^2\\
  &  & - 2 r_1 (r_3) r_2 (r_3) r_3 \cos \varphi,
\end{eqnarray*}
which can be rescaled in the same fashion as the dynamical system to obtain
\begin{eqnarray}
  E (\varphi, x) & = & \overline{\omega_{}} x^2 / 2 - (1 - x^2)^{1 / 2} \times
  \nonumber\\
  &  &  (\kappa^2 - x^2)^{1 / 2} x \cos \varphi,  \label{eq.const_E}
\end{eqnarray}
where the zero of $E$ was chosen appropriately to eliminate the dependence on
non-essential parameters.

Notice also that, the dynamical system
(\ref{eq.2d_system_x}-\ref{eq.2d_system_phi}) can be obtained from the
Hamilton equations, $\dot{\varphi} = \partial E / \partial x$ and $\dot{x} = -
\partial E / \partial \varphi$, where (\ref{eq.const_E}) is the Hamiltonian
function, and $\{ \varphi, x \}$ is the coordinate - momentum pair
respectively. This implies that the set of transformations that led from
(\ref{eq.universal_triad}) to (\ref{eq.2d_system_x}-\ref{eq.2d_system_phi})
preserve the Hamiltonian structure, i.e they were canonical transformations
[].

Since $E (\varphi (\tau), x (\tau))$ is constant along the solutions of the
dynamical system, the level sets of $E (x, \varphi)$ can be used to illustrate
the nature of the solutions.

In Fig. \ref{fig.contours_triad}, we divided the phase space $\varphi - x$, in
three sub-domains depending on the type of behavior of $\varphi$. Two regions
with bounded oscillations in $\varphi$ (in white), where $E <
\overline{\omega}$ and $E > 0$, separated by a region with unbounded evolution
of $\varphi$ (in gray), where $\overline{\omega} < E < 0$. The evolution of
$x$ is always bounded and periodic, which is expected due to $r_3 < I_{\min}$,
and because of the dimension of the space, except for the infinite time orbits
in the separatrices that only arrive (or depart) asymptotically to (from) the
saddle points. Expectedly, the frequency of the solutions will go to zero at
the separatrices and have finite values otherwise, tending to the linear
frequencies near the centers $C_0, C_{\pi}$ for close small amplitude
oscillations.

To summarize, in this particular situation, for $E < \overline{\omega} < 0$,
the amplitude of the third wave $r_3 \propto x$ oscillates around a large
value, and its phase oscillates about $\varphi = 0$. As we increase the
{\tmem{energy}}, to $\overline{\omega} < E < 0$, the fluctuations in the wave
amplitude grow, but now occur about an intermediate value, and the mismatch
phase $\varphi$, moves counterclockwise and unbounded. Finally, if we increase
the {\tmem{energy}} to $E > 0$, the fluctuations in the amplitude decrease and
occur about an even lower value, while the mismatch phase gets confined again,
but oscillates about $\varphi = \pi$. As will be illustrated now, this
analysis reverses when $\overline{\omega} > 0$.

\begin{figure}[h]
  \resizebox{210pt}{200pt}{\includegraphics{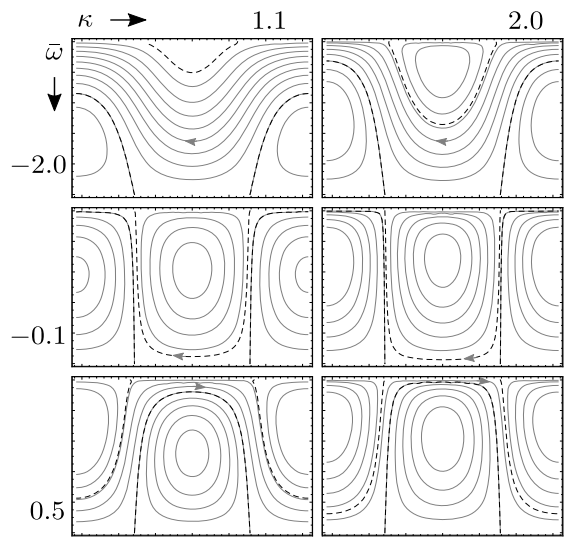}}
  \caption{\label{fig.triad_params}Phase space dependence on the control
  parameters. Large magnitudes of $\overline{\omega}$ increase the unbounded
  solutions, while large values of $\kappa$ reduce the amount of bounded
  solutions.}
\end{figure}

In Fig. \ref{fig.triad_params} we depict in few instances of phase space for
different parameters $\overline{\omega}$ and $\kappa$. The centers $C_0$ and
$C_{\pi}$, $x$-coordinate interchange as $\overline{\omega}$ goes from
negative to positive, and the separatrices flip the bounded region when the
phase mismatch flips direction as well. Also, since the unbounded motion
occurs between $E = \overline{\omega}$ and $E = 0$, larger magnitudes of
$\overline{\omega}$ lead to larger regions of unbounded motion. Conversely,
increases in $\kappa$, push the separatrices to each other, reducing the size
of the rotation region.

\subsection{Inherent modal drift}

In this work, we are interested in the development of atmospheric patterns and
their persistence in relation to atmospheric blockings. As mentioned before,
the contribution to the flow pattern from a mode $\alpha = (m, n)$ has the
form
\[ \psi_{\alpha} \propto r_{\alpha} (t) P_n^m (\cos \theta) \cos (m_i \phi +
   \varphi_i (t)), \]
where the modal phase $\varphi_i (t)$, introduces a time-dependent offset in
the zonal direction for the mode contribution to the global flow pattern. In
the context of atmospheric blocking, we can expect that the relevant mode
$\psi_3$, sustains a relatively large amplitude $r_3$ for a long time, and its
phase becomes stagnant or oscillates about the blocking longitude $\varphi_b$,
which might be correlated to the topography phase $\varphi_h$, that already
sets the zero of the mode phases in (\ref{eq.universal_triad}).

An interesting candidate for the blocking states are those orbits close to the
high amplitude center at mismatch value $\varphi = 0$ (available for
$\overline{\omega} < 0$), where the amplitude of $r_3$ is stably large and
$\varphi (t)$ oscillates about zero. This, however does not imply that the
individual phases $\varphi_i (t)$ are oscillating too, but only their
mismatch. In the following we address the conditions under which the relevant
mode $\varphi_3$ tends to oscillate about zero as well.

Consider the rescaled equation for the phase of the topography mode
$\varphi_3$
\[ \begin{array}{lll}
     \frac{d \varphi_3}{d \tau} & = & \overline{\omega}_3 - A (x) \cos
     \varphi,
   \end{array} \]
where $A (x) = (1 - x^2)^{1 / 2}  (\kappa^2 - x^2)^{1 / 2} / x$, \
$\overline{\omega}_3 = \omega_3 / I_{\min}$. Now, we consider a periodic
solution near the fixed point $C_0 = (0, x_0)$, which can be approximated in
the linear region by
\begin{eqnarray*}
  x (\tau) & \approx & x_0 + a_x \cos (\omega_0 \tau)\\
  \varphi (\tau) & \approx & a_{\varphi} \sin (\omega_0 \tau)
\end{eqnarray*}
where $\omega_0$ is the linear frequency at the fixed point $C_0$, and the
fluctuation amplitudes $a_x$, $a_{\varphi}$, satisfy $| a_x | \ll x_0$, and $|
a_{\varphi} | \ll 1$. To a second order, the evolution of the phase velocity
along the linear solution is
\begin{eqnarray*}
  \frac{d \varphi_3}{d \tau} & \approx & \overline{\omega}_3 - A (x_0) - a_x
  A' (x_0) \cos (\omega_0 \tau)\\
  &  & - \frac{a_x^2}{2} A'' (x_0) \cos^2  (\omega_0 \tau) + a^2_{\varphi} A
  (x_0) \sin^2 (\omega_0 \tau),
\end{eqnarray*}
and the drift velocity of the third phase is the average of $\dot{\varphi_3}$
taken along the orbit
\[ u_3 = \frac{1}{T} \int_0^T \frac{d \varphi_3}{d \tau} d \tau, \]
which leads to the simple form
\begin{equation}
  u_3 = \overline{\omega}_3 - A_0 + (a_{\varphi}^2 A_0 - a_x^2 A_0'') / 4,
  \label{eq.modal_drift}
\end{equation}
giving the actual drift velocity of a wave pattern in the zonal direction as a
function of the amplitude of the phase fluctuation $a_{\varphi}$ and the
coordinate $x_0$, of the fixed point $C_0$ in phase space $\varphi - x$.

In principle, the zonal drift of $\psi_3$ (\ref{eq.modal_drift}), can vanish
for a particular choice of system parameters, and at one single particular
solution in phase space, e.g., if we take an oscillation in $x - \varphi$ for
which $\overline{\omega}_3 = A_0 - (a_{\varphi}^2 A_0 - a_x^2 A_0'') / 4$, so
that the corresponding level set in $\varphi - x$ becomes the limit between
positive and negative drifting evolution for $\varphi_3$.

\begin{figure}[h]
  \resizebox{200pt}{280pt}{\includegraphics{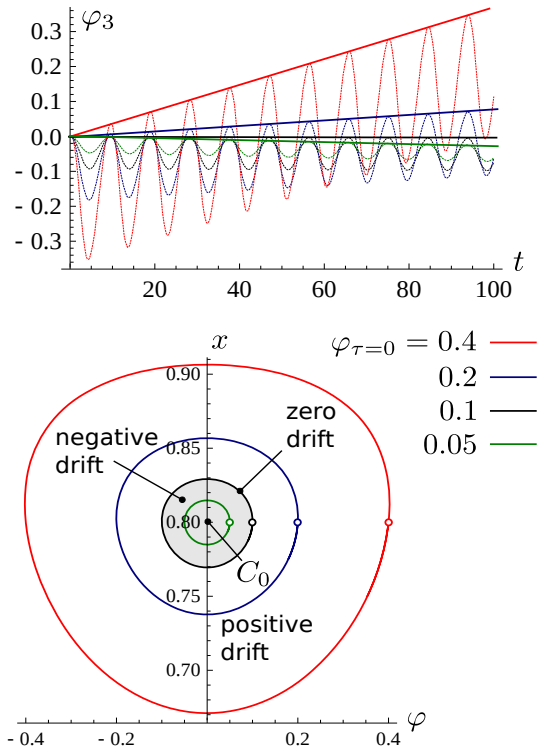}}
  \caption{\label{fig.no_drift}Dynamics of the third phase $\varphi_3$, for a
  triad supporting a zero-drift solution (black waveform and contour) at a
  stationary, high-amplitude of the third wave. At the stagnant amplitude
  $x_0$, the phase may drift in a positive or negative direction, depending on
  the initial mismatch $\varphi_0$. The zero drift solution starting at
  $(\varphi_0, x_0) = (0.8, 0.2)$ is in a closed contour in $\varphi - x$,
  about the elliptic point $C_0$, and separates positive from negative drift
  solutions for the third wave.}
\end{figure}

In Fig. \ref{fig.no_drift}, we show a particular triplet that supports
zero-drift solutions for $\varphi_3$, corresponding parameters are
$\overline{\omega} = - 1.18$, $\kappa = 1.66$, and $x_0 = 0.8$. As described
above, the zero drift solution occurs at a single level set in phase space,
and other solutions exhibit positive or negative drift depending on whether
they are inside or outside such contour.

Since zero-drift solutions of (\ref{eq.modal_drift}), have {\tmem{zero}}
measure in phase space, it is clear that the {\tmem{mode locking}} required
for the atmospheric blocking is not a statistically representative in the
space of solutions, and consequently, is not a feature on the dynamics of a
single triad, and, in general, all modes will drift with finite velocity in
the real space. In the following, we will show that mode locking is a more
general feature of the dynamics when a topography mode interacts with the
triad at particular system parameters, i.e. there is a nonzero measure region
in phase space where this phenomenon occurs.

\section{single triad with topography}\label{sec.triad_topography}

Now that we identified the fundamental parameters of a single triad without
topography, we can characterize better the effects of the interaction with a
single topography mode congruent with the third wave pattern. Before rewriting
the original system (\ref{eq.universal_triad}) in polar form, it can be seen
that $I_1$ and $I_2$ as defined in (\ref{eq.I1}, \ref{eq.I2}) are no longer
constants of motion, but we can use them to derive a new invariant of the form
\begin{equation}
  M^2 = \pm [\varepsilon_2  (| z_1 |^2 + | z_3 |^2) - \varepsilon_1  (| z_2
  |^2 + | z_3 |^2)], \label{eq.M}
\end{equation}
where the sign is chosen in such a way that $M > 0$. Provided that $M$ is a
constant, the largest Manley-Rowe quantity (mult. by the corresponding
$\varepsilon$), will be always be so, maintaining its distance to the other
quantity. In general, we can put $r_3$ in terms of $r_1, r_2$ using
(\ref{eq.M}) and the system parameters, then, the original system
(\ref{eq.universal_triad}) can be recasted in polar form to obtain the
four-dimensional system
\begin{eqnarray*}
  \dot{r_1} & = & r_2 r_3 (r_1, r_2) \sin \varphi + \varepsilon_1 r_2 \sin
  (\varphi - \varphi_3),\\
  \dot{r_2} & = & r_1 r_3 (r_1, r_2) \sin \varphi + \varepsilon_2 r_1 \sin
  (\varphi - \varphi_3),\\
  \dot{\varphi} & = & \omega + \left[ \left( \frac{r_2}{r_1} + \frac{r_1}{r_2}
  \right) r_3 (r_1, r_2) - \frac{r_1 r_2}{r_3 (r_1, r_2)} \right] \cos
  \varphi\\
  &  & + \left( \varepsilon_1  \frac{r_2}{r_1} + \varepsilon_2 
  \frac{r_1}{r_2} \right) \cos (\varphi - \varphi_3),\\
  \dot{\varphi}_3 & = & \omega_3 - \frac{r_1 r_2}{r_3 (r_1, r_2)} \cos
  \varphi,
\end{eqnarray*}
where the function $r_3$, takes an appropriate form, for instance $r_3 (r_1,
r_2) = [(M^2 + \varepsilon_1 r_2^2 - \varepsilon_2 r_1^2) / (\varepsilon_2 -
\varepsilon_1)]^{1 / 2} .$ In the case without topography the evolution was
entirely determined by the initial phase mismatch $\varphi = \varphi_3 -
\varphi_2 - \varphi_1$, and the modal amplitudes, but in this case the initial
phase of $\varphi_3$, must be specified as well.

To understand the effect of the topography in the third phase, notice that the
perturbation does not affect the equation for $\dot{\varphi_3}$ directly but
only trough its cumulative effects in $r_1 (t)$ and $r_2 (t)$, which are no
longer specified uniquely by the value of $r_3$ and the constants of motion.
As before, consider the average of the equation for $\dot{\varphi}_3$ along a
trajectory close to $C_0$.
\[ u_3 = \omega_3 - \frac{1}{T} \int_0^T \frac{r_1 (t) r_2 (t)}{r_3 (r_1 (t),
   r_2 (t))} \cos \varphi (t) \tmop{dt}, \]
where $T$ is the period of the corresponding unperturbed orbit around $C_0$.
Provided that $r_1 (t)$, $r_2 (t)$ and $\varphi (t)$ are no longer exactly
periodic, the resulting drift velocity for one cycle will differ slightly from
the next, and the drift velocity becomes a function of time.

\begin{figure}[h]
  \resizebox{200pt}{150pt}{\includegraphics{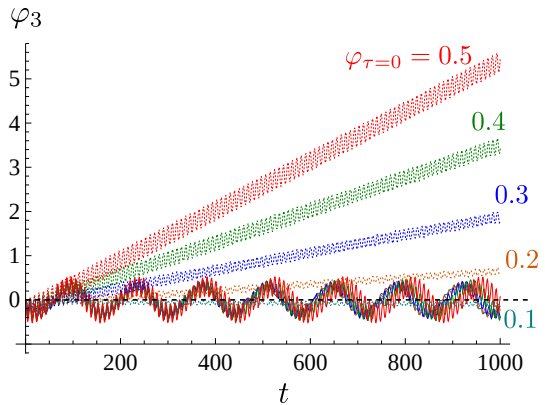}}
  \caption{\label{fig.drift_evolution}Evolution of the third phase for a
  single triad (dotted) and a triad with topography (continuous), for the same
  initial value of the amplitudes different values of the initial mismatch
  $\varphi (\tau = 0) .$}
\end{figure}

In Fig. \ref{fig.drift_evolution} we show the results of the numerical
integration of the equations of motion for one triad with $\overline{\omega} =
- 1.0, \kappa = 1.34$, with and without topography for different initial
conditions in $r_3 - \varphi$. For this choice of parameters the system
contains an elliptic fixed point with large amplitude, $x_0 = r_{3, 0} /
I_{\min} = 0.8$, around which we will study the motion with perturbation
amplitudes of $\varepsilon_1 = 0.01, \varepsilon_2 = 0.0175$. As mentioned
above the drift velocity in the unperturbed case is constant and the phase of
the third mode in general oscillates while drifting away. This occurs even
when the initial condition in phase space corresponds to bounded solutions of
the mismatch phase $\varphi$ (which is here the case). On the other hand, when
we introduce a topography mode congruent with the wave $\psi_3$, for the same
initial conditions, we obtain a bounded periodic phase drift for $\varphi_3$
about its initial value, in other words, the topography effectively
{\tmem{locks its corresponding Rossby wave spatially}} for a finite region of
phase space, i.e. in contrast with the unperturbed triad the perturbed
solutions exhibiting bounded drifts have finite measure in phase space, and
consequently finite probability when sorting initial configurations.

\begin{figure}[h]
  \resizebox{210pt}{150pt}{\includegraphics{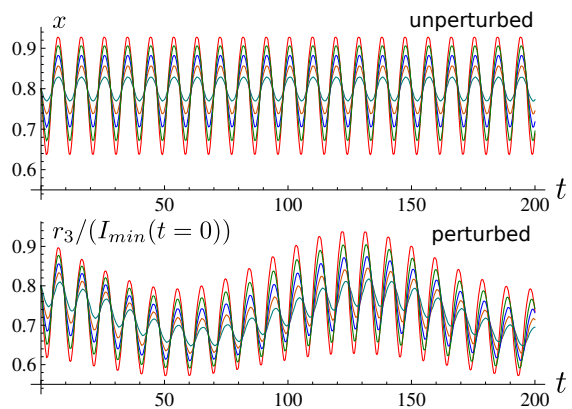}}
  \caption{\label{fig.triad_topography_nodrift} Comparison between the corresponding third mode amplitudes for
  different initial mismatches (same as Fig. \ref{fig.drift_evolution}) in the
  perturbed and unperturbed situations.}
\end{figure}

A complementary important feature of the perturbed motion is that it does not
affect importantly the amplitude of the relevant wave, but only modulates it
slightly in time. With this, we combine the two required features for
simplified atmospheric blocking, a bounded drift of the third Rossby wave with
a sustained large amplitude.

\section{conclusions}\label{sec.conclusions}

It this work we have shown that individual triads of the barotropic
quasi-geostrophic model present unbounded phase drifts on each mode, implying
that the underlying flow pattern of each wave oscillates about a rotating
phase in the real space. The drift motion is even present when the triad
mismatch phase performs bounded oscillations in phase space at high amplitudes
of a relevant mode. Particular solutions without drift do exist, but these
have {\tmem{zero}} measure in phase space. Not surprisingly, this implies that
a single triad is unable to represent the meridional persistence of
atmospheric blocking, one its most basic features. This scenario changes
drastically by the introduction of a single topography mode consistent with
one wave in the Rossby-Hawritz triad. The topography here acts as a small
perturbation on the Hamiltonian system leading to periodic changes in the
drift velocity, resulting in the non-linear phase capture of the corresponding
mode. This generates a finite volume of solutions in phase space without net
phase drift for the mode consistent with the topography, i.e. persistent
patterns have a finite probability when initial conditions are randomly
selected in phase space. Fortunately, such important changes in the phase
dynamics do not affect drastically the characteristic large amplitude of the
mode corresponding to the topographic pattern, and, consequently, the
topography-related mode can be large while its phase is bounded meridionally
in real space. The results obtained in this work are very general, as the
mechanisms mentioned here were not obtained for particular modal structures,
numerical examples were used for illustrative purposes, but they do not
restrict our analysis, which is in fact, based in a rescaled dimensionless
dynamical system encompasing an infinite family of triads. With this we are
implying that the non-linear phase capture is a common mechanism triggered by
topography (or any inhomogeneity in the atmospheric layer), and it leads to
dynamical features consistent with atmospheric blockings. The emergence of
such states or their duration might involve the interaction with other triads,
and will be investigated in a future paper, but the prescence of persistent
patterns in the elementary triads is already expected from this reductionist
treatment.

\section{Appendix}\label{sec.appendix}

\subsection{Derivation of the quasi-geostrophic
model}\label{sec.ap_model_derivation}

In general, the conservation of the potential vorticity in an incompressible
two-dimensional flow in a rotating frame with angular velocity $\Omega$ is
given by
\begin{equation}
  D_t \left( \frac{\zeta + f}{H} \right) = 0,
  \label{eq.conservation_potential-vorticity}
\end{equation}
where $\zeta = \hat{n} \cdummy \nabla \times \vec{u}$ is the relative
vorticity, with $\hat{n}$ the normal to the planet surface, $f = 2 \Omega \cos
\theta$ the planetary vorticity, and $H (\theta, \phi, t)$ the fluid depth at
the latitude $\theta$, longitude $\phi$ and time $t$. The material derivative
is defined in terms of the relative velocity in the rotating frame $D_t =
\partial / \partial t + \vec{u} \cdummy \nabla$, so that mass conservation
reads
\begin{equation}
  D_t H = - H \nabla \cdummy \vec{u}
\end{equation}
which inserted in (\ref{eq.conservation_potential-vorticity}) leads to
\begin{equation}
  D_t (\zeta + f) + (\zeta + f) \nabla \cdummy \vec{u} = 0,
\end{equation}
which can be recasted in conservation form as
\begin{equation}
  \partial_t \zeta + \nabla \cdummy [(\zeta + f)  \vec{u}] = 0,
  \label{eq.vorticity_conservation}
\end{equation}
where it becomes clear that the potential vorticity is carried by the velocity
flow.

Now, in order to obtain a useful yet simple model of the atmosphere we
consider the particular situation in which the dept $H$ is not changing in
time, so that the flow $H \vec{u}$ becomes incompressible, and can be written
in terms of a suitable stream function $\xi$ in the form
\begin{equation}
  H \vec{u} = \hat{n} \times \nabla \xi . \label{eq.mass_flow}
\end{equation}
Decomposing the depth in a large mean $\bar{H}$ and a small fluctuation
$\tilde{H}$, and dividing both sides of (\ref{eq.mass_flow}) by the mean depth
$\bar{H}$, we can write the non-solenoidal velocity field as
\begin{equation}
  \vec{u} = \frac{\hat{n} \times \nabla \psi}{(1 + \eta)},
\end{equation}
where $\eta = \tilde{H} (\theta, \phi) / \bar{H}$, and $\psi = \xi / \bar{H}$.
Replacing this form of the velocity in (\ref{eq.vorticity_conservation}) we
obtain a new form of the conservation of the absolute vorticity
\begin{equation}
  \frac{\partial \zeta}{\partial t} + J \left( \psi, \frac{\zeta + f}{1 +
  \eta} \right) = 0, \label{eq.absolute_vorticity}
\end{equation}
where $J (f_1, f_2) = \hat{n} \cdummy (\nabla f_1 \times \nabla f_2)$. This
equation, now in terms of scalar fields alone requires an additional relation
between the stream $\psi$ and the relative vorticity $\zeta$, given by
\begin{equation}
  \zeta = \frac{\nabla^2 \psi}{1 + \eta} - \frac{\nabla \eta \cdummy \nabla
  \psi}{(1 + \eta)^2}, \label{eq.vorticity_stream}
\end{equation}
which, formally, allow us to determine the evolution of the stream, provided
that we know the static relative fluctuation $\eta$. The fluctuation is due to
both the topography and the free surface, but here we will refer to it as the
topography.

\subsubsection{Perturbative expansion and hybrid model}

In the limit without topography $\eta \rightarrow 0$, the velocity field
becomes solenoidal and (\ref{eq.absolute_vorticity}) becomes the regular
non-linear form of the barotropic vorticity (BV) equation. This indicates a
functional dependence between the stream $\psi$ and the topography $\eta$,
which can be weighted by a tunable parameter $\varepsilon$, then perform a
perturbative expansion of the form
\[ \psi = \psi_0 + \varepsilon \psi_1 + \cdots, \]
The leading orders of (\ref{eq.vorticity_stream}) are
\begin{equation}
  \zeta_0 + \varepsilon \zeta_1 = \nabla^2 \psi_0 + \varepsilon (\nabla^2
  \psi_1 - \eta \nabla^2 \psi_0 - \nabla \eta \cdummy \nabla \psi_0),
  \label{eq.vorticity_expansion}
\end{equation}
which replaced in (\ref{eq.absolute_vorticity}) leads to
\begin{equation}
  \frac{\partial}{\partial t} (\zeta_0 + \varepsilon \zeta_1) + J [\psi_0 +
  \varepsilon \psi_1, (\zeta_0 + \varepsilon \zeta_1 + f) (1 - \varepsilon
  \eta)] = 0,
\end{equation}
where $\zeta_0 = \nabla^2 \psi_0$, and $\zeta_1 = \nabla^2 \psi_1 - \eta
\nabla^2 \psi_0 - \nabla \eta \cdummy \nabla \psi_0$, are the zero and first
order vorticity. At this point, if we perform a regular separation in powers
of $\varepsilon$, we recover the regular barotropic vorticity equation at
order zero and get a complicated expression for the first order, which
includes the topography. Instead of proceeding in this fashion, we consider
the influence of the topography on the zeroth order atmospheric patterns and
collect the remaining terms in an auxiliary equation indented to correct the
stream to a first order:
\begin{eqnarray}
  \frac{\partial \zeta_0}{\partial t} + J (\psi_0, \zeta_0 + f - \eta f) & = &
  0,  \label{eq.zeroth-order}\\
  \frac{\partial \zeta_1}{\partial t} + J (\psi_1, \zeta_0 + f) + J (\psi_0,
  \zeta_1 - \eta \zeta_0) & = & 0, 
\end{eqnarray}
Equation (\ref{eq.zeroth-order}) resembles the usual BV equation but now
includes a topographic forcing weighted by the Coriolis parameter $h = \eta f
= 2 \Omega \tilde{H} \cos \theta / \bar{H}$. Dropping the indices we obtain a
hybrid model between zero and one used along the text.
\begin{equation}
  \frac{\partial \zeta}{\partial t} + J (\psi, \zeta + f - h),
  \label{eq.vorticity_topography}
\end{equation}
where $\zeta = \nabla^2 \psi$, and the velocity is non-solenoidal because of
the topography correction.
\begin{equation}
  \vec{u} = (1 - h / f)  \hat{n} \times \nabla \psi .
\end{equation}
\subsection{Reduced dynamical system and modal
interaction}\label{sec.ap_reduced_system}

In general, by introduzing the stream expansion (\ref{eq.stream_expansion}),
and the topography decomposition (\ref{eq.topography_expansion}), in the
barotropic quasi-geostrophic equation (\ref{eq.dimensionless_model}). And
using the orthogonality of the spherical harmonics, it can be shown that the
dynamics of the amplitudes is given by
\begin{equation}
  k_{\gamma}  \dot{a}_{\gamma} = i m_{\gamma} a_{\gamma} + i \sum_{\alpha,
  \beta} K_{\gamma \beta \alpha} [\lambda k_{\beta} a_{\alpha} a_{\beta} +
  \varepsilon a_{\alpha} h_{\beta}],
\end{equation}
where $(\alpha, \beta, \gamma)$ are collective indices of the form $\gamma =
(m_{\gamma}, n_{\gamma})$, and $k_{\gamma} = n_{\gamma}  (n_{\gamma} + 1)_{}$.
The products between amplitudes are a consequence of the nonlinearity of the
Jacobian function, and are weighted by the interaction coefficients $K$
defined by
\begin{equation}
  K_{\gamma \beta \alpha}^{} = \int_{- 1}^1 P_{\gamma} \left( m_{\beta}
  P_{\beta} \frac{d_{} P_{\alpha}}{d z} - m_{\alpha} P_{\alpha} \frac{d_{}
  P_{\beta}}{d z} \right) d z,
\end{equation}
where $P_{\gamma} (z) = P_n^m (\cos \theta)$ are the associated Legendre
polynomial, and it can be shown that the $K$'s vanish, {\tmem{unless}} the
following conditions are satisfied.
\begin{eqnarray*}
  m_{\alpha} + m_{\beta} & = & m_{\gamma},\\
  m^2_{\alpha} + m^2_{\beta} & \neq & 0,\\
  n_{\alpha} n_{\beta} n_{\gamma} & \neq & 0,\\
  n_{\alpha} & \neq & n_{\beta},\\
  n_{\alpha} + n_{\beta} + n_{\gamma} & \tmop{is} & \tmop{odd},\\
  (n_{\alpha} - | m_a |)^2 + (n_{\beta} - | m_{\beta} |)^2 & \neq & 0,\\
  | n_{\alpha} - n_{\beta} | < & n_{\gamma} & < n_{\alpha} + n_{\beta},\\
  \beta \neq \bar{\gamma} & \tmop{and} & \alpha \neq \bar{\gamma},
\end{eqnarray*}
where $\bar{\gamma} = (- m_{\gamma}, n_{\gamma})$, for $\gamma = (m_{\gamma},
n_{\gamma})$.

\begin{eqnarray*}
  i n_1 (n_1 + 1) \dot{a}_1 & = & - m_1 a_1 + \lambda K \mu_{32} a_2^{\ast}
  a_3 \\
  &  & - \varepsilon K (h_2^{\ast} a_3 - h_3 a_2^{\ast}),\\
  i n_2 (n_2 + 1) \dot{a}_2 & = & - m_2 a_2 + \lambda K \mu_{13} a_1^{\ast}
  a_3 \\
  &  & + \varepsilon K (h_1^{\ast} a_3 - h_3 a_1^{\ast}),\\
  i n_3 (n_3 + 1) \dot{a}_3 & = & - m_3 a_3 + \lambda K \mu_{12} a_1 a_2\\
  &  & - \varepsilon K (h_2 a_1 - h_1 a_2) .
\end{eqnarray*}
\subsection{Fixed points of a single triad}\label{sec.ap_fixed_points}

In the following we classify the fixed points and stability of the single
triad two-dimensional system (\ref{eq.2d_system_x}-\ref{eq.2d_system_phi}). By
requiring $d x / d \tau = 0$, and recalling that $x \leqslant 1$, and $\kappa
\geqslant 1$ we obtain the conditions
\begin{equation}
  x_1 = 1, \quad \varphi_2 = 0, \quad \varphi_3 = \pi,
\end{equation}
Now we examine these conditions respect to the equation for the evolution in
$\varphi$
\begin{eqnarray*}
  \frac{d \varphi}{d \tau} & = & \overline{\omega} + \left( \frac{x}{\kappa^2
  - x^2} + \frac{x}{1 - x^2} - \frac{1}{x} \right) \times\\
  &  & (1 - x^2)^{1 / 2}  (\kappa^2 - x^2)^{1 / 2} \cos \varphi
\end{eqnarray*}
which, when $x \rightarrow x_1$=1 takes the form
\begin{eqnarray*}
  \frac{d \varphi}{d \tau} & \rightarrow & \overline{\omega} + \frac{(\kappa^2
  - 1)^{1 / 2}}{(1 - x^2)^{1 / 2}} \cos \varphi,
\end{eqnarray*}
so that as $x \rightarrow 1$ the motion in phase space becomes mostly
horizontal, but because of $\cos \varphi$, the direction of $d \varphi / d
\tau$ reverses at $\varphi \pm \pi / 2$, regardless of the finite offset
caused by $\overline{\omega}$, then, we have two fixed points
\begin{equation}
  x_1 = 1, \quad \varphi_1^+ = \pi / 2, \quad \varphi_1^- = - \pi / 2.
\end{equation}
Notice also, that $d \varphi / d \tau$ also diverges as $x \rightarrow 0$,
which means that, even in $d x / d \tau$ is finite, the flow in phase space
becomes horizontal, but, by the previous argument, reverses at $\varphi \pm
\pi / 2$, leading to a couple of new fixed points
\[ x_4 = 0, \quad \varphi_4^+ = \pi / 2, \quad \varphi_4^- = - \pi / 2. \]
For the remaining fixed points we insert $\varphi_2$ and $\varphi_3$ in $d
\varphi / d \tau = 0$, to obtain
\begin{equation}
  \overline{\omega} x \sqrt{(1 - x^2) (\kappa^2 - x^2)} \pm [2 (1 + \kappa^2)
  x^2 - 3 x^4 - \kappa^2] = 0 \label{eq.dot_phi_0}
\end{equation}
which can be written as quartic polynomial equation in $x^2$, and can be
solved using computational algebra, but such solution is extensive and has
little value for phenomenological interpretation. However, we can guarantee
that such solution exist for both $\varphi_2$ and $\varphi_3$. To see this,
note that the l.h.s. function in (\ref{eq.dot_phi_0}) can be written as
\[ f_{\tmop{lhs}} (x) = \overline{\omega} f_1 (x) \pm f_2 (x), \]
where $f_1 (0) = f_1 (1) = 0$, and $f_2 (0) = - \kappa^2$, $f_2 (1) = \kappa^2
- 1$. Provided that $f_2 (0) < 0$ and $f_2 (1) > 0$, the function
$f_{\tmop{lhs}}$ changes sign between $x = 0$ and $x = 1$, \ regardless of the
sign of $\overline{\omega}$. Given that $f_{\tmop{lhs}} (x)$ has no
singularities it follows that it must vanish {\tmem{at least}} one time in $0
< x < 1$. With this we guarantee that there is at least one pair of fixed
point solutions
\[ \varphi_2 = 0 \quad, 0 < x_2 < 1, \quad \varphi_3 = \pi, \quad 0 < x_3 < 1,
\]
with further analysis we can make the intervals narrower, but here we are only
interested in the existence of such fixed points. To summarize we have
\begin{eqnarray*}
  S_1^{\pm} : \quad (\varphi_1^{\pm}, x_1) & = & (\pm \pi / 2, 1) \rightarrow
  \tmop{hyperbolic},\\
  S_0^{\pm} : \quad (\varphi_1^{\pm}, x_4) & = & (\pm \pi / 2, 0) \rightarrow
  \tmop{hyperbolic},\\
  C_0 : \quad (\varphi_2, x_2) & = & (0, 0 < x_2 < 1) \rightarrow
  \tmop{Elliptic},\\
  C_{\pi} : \quad (\varphi_3, x_3) & = & (\pi, 0 < x_2 < 1) \rightarrow
  \tmop{Elliptic} .
\end{eqnarray*}
The classification of these equilibria was made through the discriminant of
the Jacobian matrix, and it wont be explicited here.

\end{document}